\documentclass[12pt]{iopart}

\usepackage{graphicx}

\begin{document}

\title{Gate dependent Raman spectroscopy of graphene on hexagonal boron nitride}

\author{Kanokporn Chattrakun$^1$, Shengqiang Huang$^1$, K. Watanabe$^2$, T. Taniguchi$^2$, A. Sandhu$^{1,3}$ and B. J. LeRoy$^1$}
\address{$^1$ Department of Physics, University of Arizona, Tucson, AZ, 85721 USA.}
\address{$^2$ National Institute for Materials Science, 1-1 Namiki, Tsukuba, 305-0044 Japan.}
\address{$^3$ College of Optical Sciences, University of Arizona, Tucson, AZ, 85721 USA.}
\ead{leroy@physics.arizona.edu}

\begin{abstract}
Raman spectroscopy, a fast and nondestructive imaging method, can be used to monitor the doping level in graphene devices.  We fabricated chemical vapor deposition (CVD) grown graphene on atomically flat hexagonal boron nitride (hBN) flakes and SiO$_2$ substrates.  We compared their Raman response as a function of charge carrier density using an ion gel as a top gate.  The G peak position, 2D peak position, 2D peak width and the ratio of the 2D peak area to the G peak area show a dependence on carrier density that differs for hBN compared to SiO$_2$. Histograms of two-dimensional mapping are used to compare the fluctuations in the Raman peak properties between the two substrates. The hBN substrate has been found to produce fewer fluctuations at the same charge density owing to its atomically flat surface and reduced charged impurities.
\end{abstract}


\maketitle
One of the major limits for graphene device performance is the presence of electron and hole puddles caused by charged impurities in the SiO$_2$ substrate and its roughness.  Using hexagonal boron nitride (hBN) as a substrate has lead to significantly increased mobility~\cite{Dean2010,Kim2011} and drastically reduced electron and hole puddles as compared to SiO$_2$~\cite{Xue2011,Decker2011} for both exfoliated and chemical vapor deposition (CVD) grown graphene.  These improvements are due to the atomic flatness of the hBN substrate and the reduction in charged impurities.  Having an improved substrate for graphene presents an opportunity to fabricate high mobility nanoelectronic devices. There are several approaches to obtain a single layer graphene sheet. One of which is the CVD method which can produce monolayer areas that are single domain over areas as large as 20 $\mu$m~\cite{Reina2009,Li2011}.  Using the CVD growth method gives an advantage over other methods such as mechanical exfoliation which produces considerably smaller graphene flakes and is therefore not scalable for large area applications. Moreover, since we would like to compare results between two substrates, hBN and SiO$_2$, we want to use large domains of graphene so that an individual domain will be on both substrates. 

Raman spectroscopy is a fast and nondestructive imaging method that can be used to monitor many properties of graphene such as the number of layers~\cite{Ferrari2006}, defect density~\cite{Cancado2011}, impurity density~\cite{Casiraghi2007}, and doping level~\cite{Das2008}. There are two prominent peaks in the Raman spectrum of high quality monolayer graphene; the G peak and the 2D peak. The G peak originates from a first order scattering process involving a doubly degenerate phonon, E$_{2g}$, at the {\bf $\Gamma$} point, the center of the graphene Brillouin zone.  The 2D peak originates from a second order, double resonance process consisting of two phonons both located near the {\bf K} point, the corner of the Brillouin zone.  The peak position of the 2D peak is excitation energy dependent due to the combination of the phonon dispersion relation and the requirement for energy and momentum conservation~\cite{Malard2009,Ferrari2013}.  Doping of graphene shifts its Fermi level leading to controllable changes in its electronic properties as well as the Raman peaks.  Doping can be accomplished by electrostatic backgating~\cite{Kim2011, Ponomarenko2009,Schwierz2010}, chemical methods~\cite{Wang2012a}, and electrochemical methods~\cite{Das2008}.  Electrochemical doping is able to induce a high charge density in the graphene and is therefore used in this study.  Properties such as the peak positions (frequencies), widths, and the ratio of the 2D and G intensities can be used to probe the doping level in graphene. It has been shown that the G peak width~\cite{Yan2007} and position~\cite{Das2008,Yan2007,Pisana2007} on SiO$_2$ are sensitive to the charge density. The G peak blue shifts with increasing doping of graphene and saturates at high doping due to the non-adiabatic removal of the Kohn anomaly at the {\bf $\Gamma$} point~\cite{Pisana2007}.  The G phonon lifetime quickly increases with carrier density because the excited electron cannot decay due to the Pauli exclusion principle and momentum conservation. Thus the G peak width, which depends inversely on the phonon lifetime strongly depends on the charge carrier density. The 2D peak intensity~\cite{Casiraghi2009,Basko2009}, and position~\cite{Stampfer2007} on SiO$_2$ also depend on doping. The 2D peak width does not demonstrate such sensitivity since the decay process is unaffected by the Pauli exclusion principle.  The 2D peak intensity decreases with increasing charge density due to the increasing of the electron-electron scattering rate at high density~\cite{Basko2009}.  When graphene is placed on hBN, the G peak has been reported to red-shift~\cite{Wang2012b,Forster2012,Ahn2013} while the shift of the 2D peak is reported to depend on the sample preparation procedure~\cite{Wang2012b,Ahn2013}.  The carrier dependent Raman properties of graphene have been widely and intensively studied on SiO$_2$ but the systematic doping dependence on hBN has not yet been investigated.

Here we report Raman spectroscopy measurements as a function of doping level for graphene on hBN and SiO$_2$.  Graphene was synthesized by low-pressure CVD on Cu foil~\cite{Li2011}. The foil was heated to 1040 $^{\circ}$C and annealed for 20 minutes under a H$_2$ gas flow rate of 4 sccm at a pressure of 20 mTorr. Then the Cu was exposed to CH$_4$ and H$_2$ gas at 1.3 and 4 sccm respectively for 1 hour. The system was cooled down to 350 $^{\circ}$C under the same CH$_4$ and H$_2$ flow rate.  Using large area CVD graphene allows the same graphene flake to be probed both on the hBN and on the underlying SiO$_2$.  Single crystal hBN flakes~\cite{Taniguchi2007} were mechanically exfoliated onto a Si substrate with 285 nm thick SiO$_2$.  After exfoliating the hBN, the samples were annealed under O$_2$ and Ar gas at 500 $^{\circ}$C for 3 hours to remove any excess tape residue. PMMA was spun on top of the Cu foil then an aqueous HCl solution was used to etch the Cu. The graphene on PMMA was thoroughly rinsed in deionized water and deposited on the hBN flakes. The sample was soaked in acetone to dissolve the PMMA, then rinsed in IPA and dried with ultra-high purity nitrogen gas. Thermal annealing under H$_2$ and Ar gas was performed at 350 $^{\circ}$C for 3 hours to remove any PMMA residue~\cite{Ishigami2007}.  Cr/Au contacts were written using electron beam lithography for electrical transport measurements. The device was annealed in H$_2$ and Ar gas at 350 $^{\circ}$C for another 3 hours after deposition of the metal contacts before any measurements. The doping was accomplished using electrochemical gating using a polymer electrolyte, LiClO$_4$:PEO~\cite{Das2008} (1:8 by weight). Raman spectroscopy was performed using a Nd:YAG laser (532 nm) on graphene with and without hBN underneath, the spot size was ~1 $\mu$m with a 100x (NA = 0.8) objective lens. The voltage on the ion gel was varied from -1.5 V to 3 V. Two-dimensional Raman mapping was performed at -1.5, 0, and 2 V on the ion gel in the area shown in figure 1a. We studied the G peak position, 2D peak position, 2D peak width, and the ratio of the 2D to G peak areas as a function of doping level on both the hBN flakes and on the SiO$_2$ substrate.

Figure 1(a) shows an optical image of one of the devices measured. The entire area is covered with graphene; the darker area on the left has hBN with a uniform thickness of 10.5 nm determined from its optical contrast~\cite{Golla2013}; the bright green area near the center of the image is thicker hBN; the area on the right has only graphene on SiO$_2$.  The total size of the image is 12.6 by 3.6 $\mu$m and the scale bar is 2 $\mu$m. The dashed line indicates a region of bilayer graphene (the thicker hBN and bilayer graphene regions, shown in dark grey in figures 2(a), 3(a), and 4(a), are excluded from all of our analyses).  Figure 1(b) shows a schematic diagram of the experimental setup used for the two-dimensional Raman mapping measurements. A Nd:YAG laser, wavelength of 532 nm, is sent through a 100x objective lens and focused onto the sample.  The reflected light is dispersed using a 600 lines/mm grating and the resulting spectrum is imaged on a thermoelectrically cooled CCD.  The spectral resolution is ~1 cm$^{-1}$.  For two-dimensional mapping, the sample position is moved using a closed-loop piezoelectric stage.  The Raman spectrum has two peaks from the graphene; the G peak at ~1590 cm$^{-1}$, and the 2D peak at ~2690 cm$^{-1}$~\cite{Ferrari2006,Graf2007}, one peak due to hBN at 1365 cm$^{-1}$~\cite{Geick1966} and two additional peaks from the ion gel near 1280 cm$^{-1}$ and 1480 cm$^{-1}$~\cite{Yoshihara1964}.  A typical Raman spectrum is shown in figure 1(c) showing a 2D peak that is several times stronger than the G peak.  The spectra are fitted with a single Lorentzian for all G, 2D and hBN peaks. The 2D peak can be fitted with a single Lorenztian because the graphene is supported on hBN or SiO$_2$ and the system has sufficiently high level of doping to overcome the intrinsic bimodal lineshape of the 2D peak seen in freestanding graphene~\cite{Berciaud2013}.  Figure 1(d) plots the ratio of the integrated area of the 2D and G peak for graphene on hBN (blue) and SiO$_2$ (red) as a function of the voltage applied to the ion gel. For stability of the sample, small steps and extended time between each voltage step are required. The voltage was applied in steps of 0.05 V, after each step there was a waiting time of one minute for the ion gel charge to stabilize before collecting data. The area ratio on hBN is greater than that on SiO$_2$ and both vary as a function of gate voltage or charge density. The ratio has a maximum value near the Dirac point of the sample (~-1V) and decreases as the charge density is increased on the graphene for both holes and electrons.  As the voltage on the ion gel is changed, the Fermi energy of the graphene shifts.  As the Fermi energy shifts, the intensity of the G peak remains constant but the intensity of the 2D peak is reduced leading to a decrease in the ratio of the 2D peak area to the G peak area. The 2D peak intensity depends on doping through the scattering rate of the photoexcited electrons and holes. The rate of electron-electron collisions is sensitive to the density of carriers, increasing the scattering rate as the charge density increases resulting in a decrease of the 2D peak intensity~\cite{Basko2009}.

\begin{figure}[h]
\centering
\includegraphics[width=0.75 \textwidth]{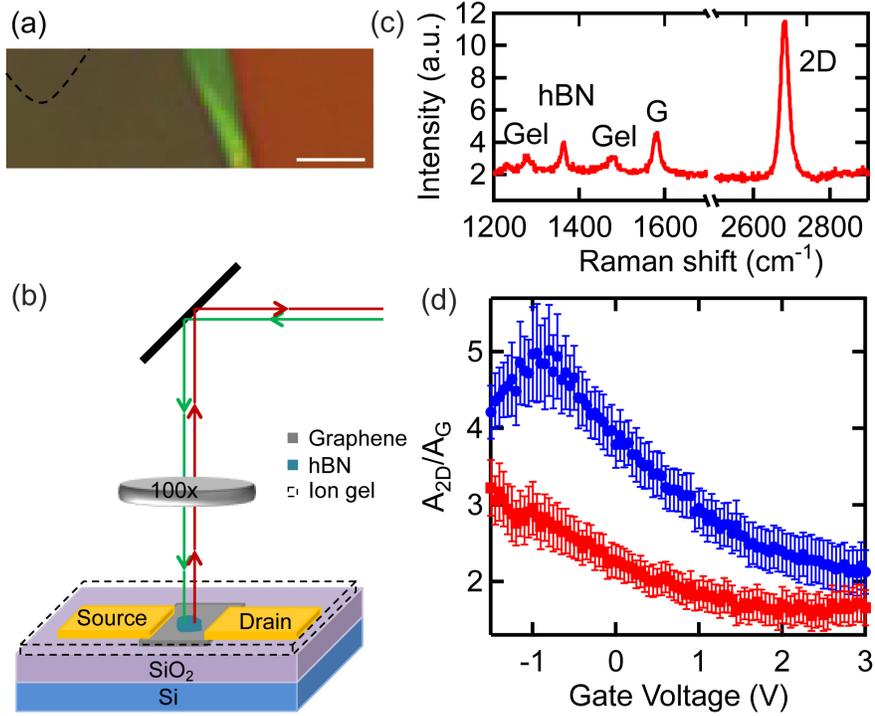}
\caption{\label{fig:schematic} (colour online) (a) Optical microscope image of graphene on hBN. The scale bar is 2$\mu$m. The dashed line in the top left corner indicates a bilayer graphene region and the bright green area indicates thicker hBN, which are both excluded in the analyses and shown as dark grey areas in figure 2(a), 3(a), and 4(a). (b) Schematic diagram of the Raman spectroscopy set up. The grey area in between the source and drain electrodes is graphene. The blue area underneath graphene is hBN. The dashed box shows that the whole device is covered with an ion gel (PEO+LiClO$_4$) serving as a top gate. (c) A typical Raman spectrum of graphene on hBN with ion gel. G and 2D denote Raman peaks from graphene. hBN and Gel denote Raman peaks from the hBN substrate and ion gel respectively. (d) Area ratio of the 2D to the G peak (A$_{\mathrm{2D}}$/A$_{\mathrm{G}}$) as a function of gate voltage applied to the ion gel on hBN (blue) and SiO$_2$ (red).}
\end{figure}

Figure 2(a) shows a two dimensional Raman map of the G peak position ($\omega_G$) for the area shown in figure 1(a).  From the map, the G peak position is lower on hBN than on SiO$_2$.  The data in figure 2(a) was acquired with 0 V on the ion gel.  The range of peak positions is 1584-1598 cm$^{-1}$ on hBN and 1588-1602 cm$^{-1}$ on SiO$_2$.  We have performed similar analysis on ten other flakes and found that the G peak is on average 3 cm$^{-1}$ lower on hBN compared to SiO$_2$.  We have found that the thickness of the hBN has a strong effect on the strength of the G and 2D peaks in graphene and therefore we have only performed the analysis for thin hBN flakes of uniform thickness.  G peak positions on both hBN (blue) and SiO$_2$ (red) shift to higher frequency for charge densities that are away from the neutrality point for both electrons and holes as shown in figure 2(b).  The increase of the G peak position along with the narrowing of the peak, which is also observed in our measurements, is a result of increased charge density~\cite{Pisana2007,Yamamoto2012}.  From the shift of the G peak position with the applied voltage, we determined the maximum induced carrier density to be $1\times10^{13}$ cm$^{-2}$ when 3 V was applied to the ion gel~\cite{Lazzeri2006,Chen2011}. From our two-dimensional mapping of ten different graphene on hBN flakes, we found that the average standard deviation of the histogram of the G peak position was 1 cm$^{-1}$ on hBN and 1.6 cm${-1}$ on SiO$_2$.  These values can be converted to fluctuations in energy by using the movement of the G peak position with energy (42 cm$^{-1}$/eV)~\cite{Chen2011}.  These fluctuations in energy are then converted to variations in charge density using $n=\frac{1}{\pi}(\frac{E_F}{\hbar v_F})^2$, where $v_F$ is the Fermi velocity which is $1.1\times10^6$ m/s in graphene, giving average charge fluctuations of $6.1\times10^{10}$ cm$^{-2}$ on hBN and $1.56\times10^{11}$ cm$^{-2}$ on SiO$_2$.  As observed in other local spectroscopy measurements, the charge fluctuations in graphene are reduced on the hBN as compared to the SiO$_2$~\cite{Xue2011,Decker2011}.  Figures 2(c), 2(d), and 2(e) are histograms extracted from the two dimensional mappings at voltages on the ion gel of -1.5, 0, and 2 V respectively. The G peak positions are lower on hBN (blue) than SiO$_2$ (red) for the 0 and 2 V gate voltages, but approximately the same at -1.5 V. The difference of the peak positions at 0 V is more pronounced than 2 V. This indicates the G peak on hBN moves more slowly than on SiO$_2$ at negative voltages and faster at positive voltages. 

\begin{figure}[h]
\centering
\includegraphics[width=0.75 \textwidth]{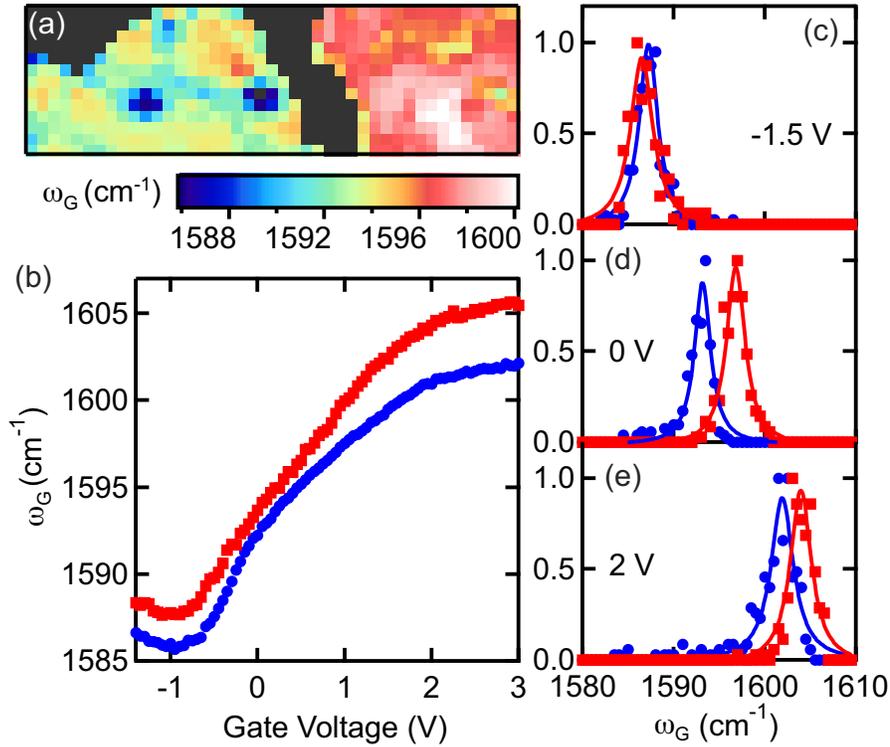}
\caption{\label{fig:GPeak} (colour online) G peak position analyses. (a) Two-dimensional Raman mapping of the G peak position ($\omega_G$) at 0 V on the ion gel. Each pixel corresponds to an area of 0.3x0.3 $\mu$m$^2$.  (b) The G peak position of graphene dependence on the voltage applied to the ion gel on hBN (blue) and SiO$_2$ (red) with the gate voltage varying from -1.5 V to 3 V. The total movement of the G peak position can be used to determine the maximum carrier density to be $2\times10^{13}$ cm$^{-2}$ when 3 V was applied to the ion gel. (c), (d), (e) Normalized histograms extracted from the two-dimensional Raman mapping of the G peak position at -1.5, 0, 2 V on the ion gel respectively. The red squares represent the histograms extracted from graphene on SiO$_2$ and the blues circles represent those from graphene on hBN.}
\end{figure}

A two-dimensional map of the 2D peak position ($\omega_{2D}$) with 2 V on the ion gel ($n = 9\times10^{12}$ cm$^{-2}$) is shown in figure 3(a). The figure indicates a significant difference in the 2D peak position on the hBN flake compared to the SiO$_2$ substrate.  The shift of the 2D peak position can be used to indicate the doping level. Figure 3(b) plots the 2D peak position as a function of doping on local spots of graphene on SiO$_2$ and hBN.  It decreases for $n$ doping and increases for $p$ doping on both substrates. At high electron concentrations, the 2D peak strongly blue-shifts on SiO$_2$ but there is only a much weaker shift on hBN.  The shift in the 2D position on hBN results from a reduction of the Kohn anomaly at the K point influenced by an enhanced screening from the dielectric substrate, which is not a direct interaction between graphene and the substrate. The electron-electron interaction depends on the electronic screening by the environment. Therefore such interaction is expected to be reduced by a dielectric substrate. Although hBN has a similar dielectric constant to SiO$_2$, its atomically flat surface provides a different dielectric environment for graphene. The reduced charged impurities and increased flatness affects the Fermi velocity of the carriers as well as the electron-electron and electron phonon coupling in graphene, thus enhancing the dielectric screening for graphene on hBN~\cite{Forster2012,Graf2007}.   Figures 3(c), 3(d), and 3(e) are histograms extracted from the two-dimensional mapping at -1.5, 0, and 2 V on the ion gel respectively.  The histograms of the 2D peak position at 0 V (figure 2(c)) are nearly identical between the two substrates; hBN (blue) and SiO$_2$ (red). For the particular location chosen for figure 3(b) the 2D peak is slightly blue shifted on the SiO$_2$ compared to the hBN.  At both positive and negative gate voltage, the 2D peak positions show more spatial variation on both substrates compared to the 0 V. The reduced movement of the 2D peak on hBN is clearly seen in figure 3(d), where the two distributions no longer overlap. 

\begin{figure}[t]
\centering
\includegraphics[width=0.75 \textwidth]{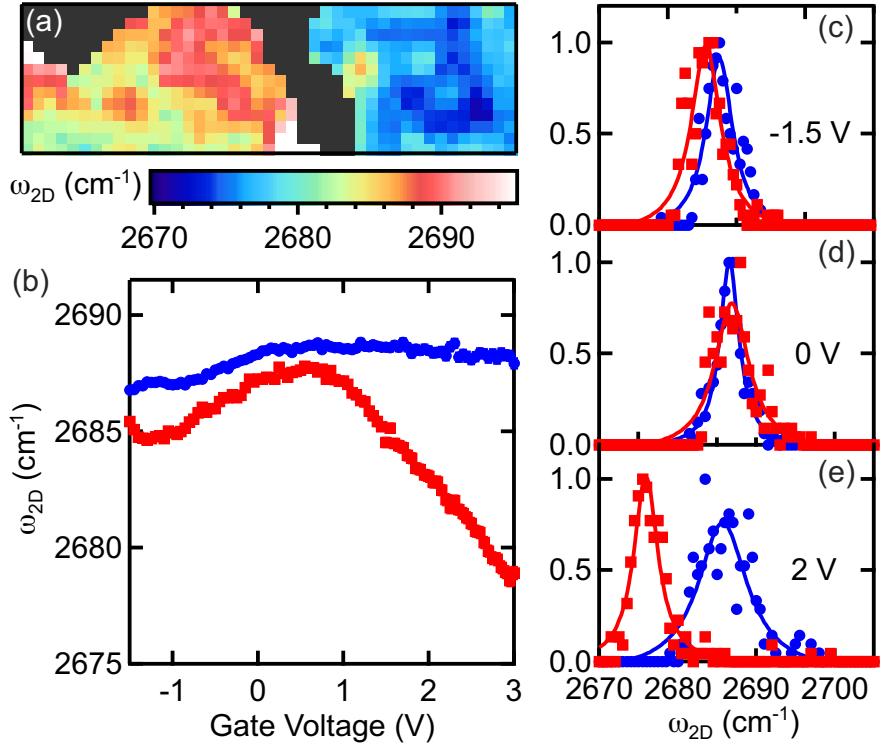}
\caption{\label{fig:2DPos} (colour online) 2D peak position analyses. (a) Two-dimensional Raman mapping of the 2D peak position ($\omega_{2D}$) with 2 V on the ion gel.  Each pixel corresponds to an area of 0.3x0.3 $\mu$m$^2$.  (b) The 2D peak position dependence on the voltage applied to the ion gel on hBN (blue) and SiO$_2$ (red) with the gate voltage varying from -1.5 V to 3 V. (c), (d), (e) Normalized histograms extracted from the two-dimensional Raman mapping of the 2D peak position at -1.5, 0, 2 V on the ion gel respectively. The red squares represent the histograms extracted from graphene on SiO$_2$ and the blues circles represent those from graphene on hBN. }
\end{figure}

Figure 4(a) maps the 2D peak width ($\Gamma_{2D}$) at 0 V on the ion gel. The map indicates the 2D peak is narrower on hBN compared to SiO$_2$.  Using hBN as a substrate for graphene, has been shown to lead to a reduction in surface roughness and charged impurities~\cite{Dean2010,Xue2011,Decker2011}.  Both of these effects reduce the phonon scattering rate and therefore lead to a longer phonon lifetime and hence a narrower 2D peak~\cite{Pisana2007,Forster2012,Graf2007}.  Figure 4(b) plots the 2D peak width as a function of the applied voltage on the ion gel. The hBN substrate is shown in blue, and the SiO$_2$ substrate is red.  Since the peak width originates from the scattering of phonons, the SiO$_2$ substrate that has greater scattering responds less to the applied gate voltage or carrier concentration because of the high scattering rate even near the Dirac point. On the other hand, the hBN substrate that has less scattering at low charge density responds faster to the increasing carrier density as seen by the increasing 2D width at higher applied voltage.  Figures 4(c), 4(d), and 4(e) are histograms of the 2D peak widths at -1.5, 0, and 2 V gate voltages respectively. They show that the graphene 2D peak is narrower on hBN than SiO$_2$ for all charge densities however the width of the 2D peak increases at higher charge density (2 V on the ion gel).  The distribution of the histogram gives a measure of the fluctuations on the two substrates.  Near the Dirac point (figure 4(c)), the widths of the distributions are comparable as the FWHM is 3.3 cm$^{-1}$ on hBN and 3.7 cm$^{-1}$ on SiO$_2$. When further away from the Dirac point (figures 4(d) and 4(e)), the hBN substrate shows a narrower distribution, indicating less charge fluctuations~\cite{Wang2012b}.  The histograms in figures 4(d) and 4(e) have FWHM of 2.7, 4.7 cm$^{-1}$ on hBN and 4.2, 6.0 cm$^{-1}$ on SiO$_2$ respectively.

\begin{figure}[t]
\centering
\includegraphics[width=0.75 \textwidth]{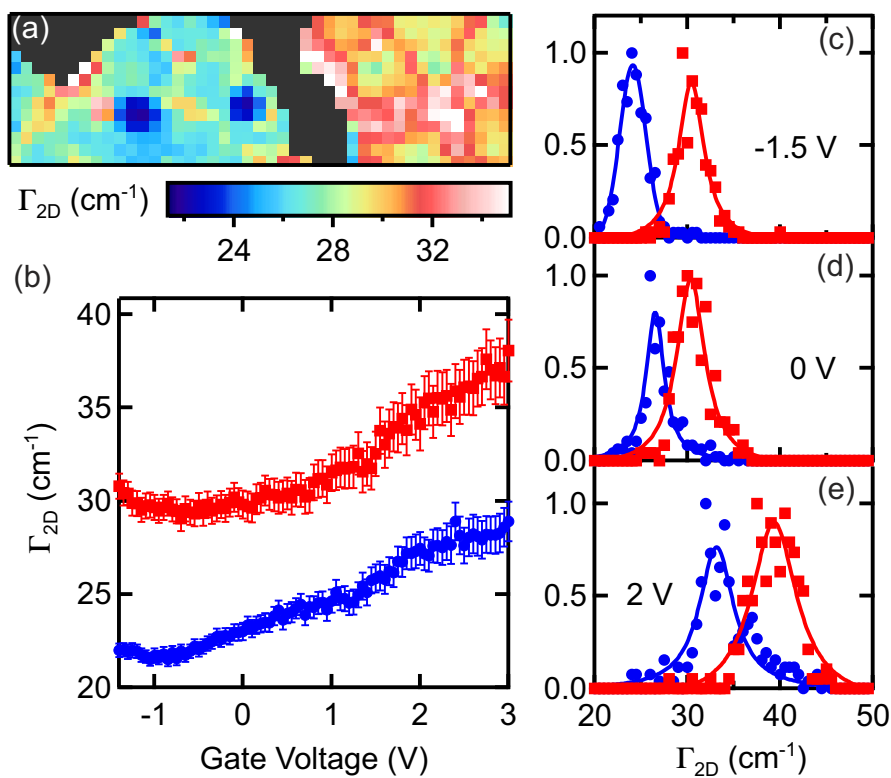}
\caption{\label{fig:2DWidth} (colour online) 2D peak width analyses. (a) Two-dimensional Raman mapping of the 2D peak width ($\Gamma_{2D}$) at 0 V on the ion gel. Each pixel corresponds to an area of 0.3x0.3 $\mu$m$^2$.  (b) The 2D peak position dependence on the voltage applied to the ion gel on hBN (blue) and SiO$_2$ (red) with the gate voltage varying from -1.5 V to 3 V (c), (d), (e) Normalized histograms extracted from the two-dimensional Raman mapping of the 2D peak width at -1.5, 0, 2 V on the ion gel respectively. The red dots represent the histograms extracted from graphene on SiO$_2$ and the blues ones represent those from graphene on hBN.}
\end{figure}

We have investigated the Raman properties of graphene such as the G peak position, the 2D peak position, the 2D peak width, and the area ratio of the 2D peak to the G peak as a function of carrier density on two different substrates; hBN and SiO$_2$. We have found that the area ratio has a higher value on hBN and varies with the carrier density. The G peak position depends on the carrier density, and the position of the G peak on hBN is lower than on SiO$_2$ for the same charge density. The analysis of two-dimensional mapping results show less fluctuation of the G peak position on hBN indicating less charge fluctuation on the hBN substrate as compared to SiO$_2$.  We have found a reduction by a factor of three in the charge fluctuation on hBN as compared to SiO$_2$.  The 2D peak positions are comparable near the Dirac point on both substrates but show a strong blue-shift at high density on SiO$_2$ while remaining approximately constant on hBN. The 2D peak width has a lower value on hBN but shows more variation at high carrier concentration indicating less scattering. The distribution of the histograms has been found to be narrower on hBN, and becomes more pronounced at higher density. This reflects a lower fluctuation of carrier density on graphene when hBN is used as a substrate.  The variation of the Raman peaks with charge density can be used to quickly identify the doping in graphene on hBN devices.

\ack
K. C., S. H. and B. J. L. acknowledge support from the U. S. Army Research Office under contract W911NF-09-1-0333 and NSF CAREER Award No. DMR/0953784.


\end{document}